\documentstyle[emulateapj]{article}
\lefthead{Hughes}
\righthead{X-Ray Expansion of Kepler's SNR}
\received{12 May 1999}
\submitted{The Astrophysical Journal: Received 12 May 1999; Accepted
11 July 1999}
\def\einstein{{\it Einstein}}
\def\EO{{\it Einstein Observatory}}
\def\rosat{{\it ROSAT}}
\def\asca{{\it ASCA}}
\def\astroe{{\it Astro-E}}
\def\xmm{{\it XMM}}
\def\chandra{{\it Chandra}}
\def\mydegree{^\circ\mskip-5mu}
\def\myarcmin{^\prime\mskip-5mu}
\def\myarcsec{\mskip1mu^{\prime\prime}\mskip-7mu.\mskip2mu}
\def\lsim{\hbox{\raise.35ex\rlap{$<$}\lower.6ex\hbox{$\sim$}\ }}
\def\gsim{\hbox{\raise.35ex\rlap{$>$}\lower.6ex\hbox{$\sim$}\ }}

\begin{document}

\title{The Extraordinarily Rapid Expansion of the X-ray Remnant of
Kepler's Supernova (SN1604)} 

\author{John P. Hughes}
\affil{Department of Physics and Astronomy, Rutgers
University, \\136 Frelinghuysen Road, Piscataway, NJ 08854-8109 \\ 
jph@physics.rutgers.edu}

\begin{abstract}

Four individual high resolution X-ray images from \rosat\ and the \EO\
have been used to measure the expansion rate of the remnant of
Kepler's supernova (SN 1604).  Highly significant measurements of the
expansion have been made for time baselines varying from 5.5 yrs to
17.5 yrs.  All measurements are consistent with a current expansion
rate averaged over the entire remnant of $0.239^{+0.015\,+0.017}
_{-0.015\,-0.010} \,\% \,\rm yr^{-1}$, which, when combined with the
known age of the remnant, determines the expansion parameter $m$,
defined as $R\propto t^m$, to be $0.93^{+0.06\,+0.07}
_{-0.06\,-0.04}$. The error bars on these results include both
statistical (first pair of errors) and systematic (second pair)
uncertainty.  According to this result the X-ray remnant is expanding
at a rate that is remarkably close to free expansion and nearly twice
as fast as the mean expansion rate of the radio remnant.  The
expansion rates as a function of radius and azimuthal angle are also
presented based on two \rosat\ images that were registered to an
accuracy better than $0\myarcsec5$.  Significant radial and azimuthal
variations that appear to arise from the motion of individual X-ray
knots are seen.  The high expansion rate of the X-ray remnant appears
to be inconsistent with currently accepted dynamical models for the
evolution of Kepler's SNR.

\end{abstract}


\section{Introduction}

The remnant of Kepler's supernova (SN) is the youngest of the
historical supernova remnants (SNRs) with an accurate,
well-established explosion date (A.D. 1604).  Although long suggested
to be the remnant of a Type I SN based on its light curve (see Doggett
\& Branch 1985), recently it has been proposed that it was the
explosion of a massive star.  Numerous studies of Kepler's SNR, from
the optical work of van den Bergh \& Kamper (1977) to the X-ray
studies of White \& Long (1983) and Hughes \& Helfand (1985), have
concluded that SN 1604 exploded in a dense environment, a conclusion
that is strongly at odds with the remnant's large distance above the
Galactic plane.  The model introduced by Bandiera (1987) in which the
progenitor was a rapidly-moving mass-losing star, provided a
significant breakthrough in our understanding of the nature of
Kepler's SNR. This model was able to explain both the high ambient
density and the strongly asymmetric appearance of the remnant in both
the X-ray and radio bands.  However, it also indicated that the
dynamics of the remnant should be extremely complex with reflected and
transmitted shocks propagating through the SN ejecta as well as
pre-existing structures in the circumstellar medium (CSM) (Borkowski,
Blondin, \& Sarazin 1992). In light of this complexity it is clear
that measurements of the dynamics of Kepler's SNR using as many
different methods as possible, are going to be an important ingredient
in understanding the evolution of the remnant.  In this article I
present the first results on the current expansion rate of the X-ray
remnant of Kepler's supernova.

\par

Previous expansion measurements of Kepler's SNR have been done in both
the optical and radio band.  For reference the time-averaged expansion
rate of Kepler's SNR based on the full extent of the remnant in either
the radio or X-ray band and its well-known age is $\sim$$0\myarcsec25$
yr$^{-1}$.  The brighter optical knots in Kepler are expanding quite
slowly, with an expansion timescale of $\sim$32,000 yr (Bandiera \&
van den Bergh 1991), a rate that is about two orders of magnitude
slower than the time-averaged expansion of the remnant itself.  Their
slow motion and high nitrogen abundance (Dennefeld 1982; Blair, Long,
\& Vancura 1991) indicate a circumstellar origin for the dense
radiative knots, an idea originally proposed by van den Bergh \&
Kamper (1977).  The discovery of nonradiative, Balmer-dominated,
optical filaments in Kepler's SNR (Fesen et al.~1989) provided the
opportunity to measure the shock velocity (albeit at a few specific
locations in the remnant) from the width of the broad H$\alpha$
emission-line, which Blair et al.~(1991) found to yield average shock
velocities of 1550--2000 km s$^{-1}$.

\begin{deluxetable}{cccc}
\tablecaption{Observations of Kepler's SNR}
\tablewidth{4.5truein}
\tablehead{
\colhead{Observatory} & \colhead{Start Date} &
\colhead{Average MJD} & \colhead{Duration (s)} 
}
\startdata
{\it Einstein} (E1) & 1979 Sep 29 & 44146.62 & $\phantom{0}5593.9$\nl
{\it Einstein} (E2) & 1981 Mar 22 & 44686.33 & $12064.0$\nl
{\it ROSAT}    (R1) & 1991 Sep 11 & 48511.17 & $36143.2$\nl
{\it ROSAT}    (R2) & 1997 Mar 10 & 50522.68 & $67895.3$\nl
\enddata
\end{deluxetable}

Radio images of Kepler's SNR from the VLA taken at two epochs (1981
and 1985) were used by Dickel et al.~(1988) to determine the current
expansion rate.  Since the age of the remnant is known, it is possible
to determine the expansion parameter, $m$, which is defined as
$R\propto t^m$ (i.e., the remnant's radius is assumed to evolve as a
power-law with age). Dickel et al.~(1988) found a best-fit value of $m
= 0.50$ averaged over the entire remnant, with a likely range
including random and systematic uncertainties of $0.49 < m < 0.56$.
Variations in the expansion rate with azimuthal angle were seen
too. The expansion parameter of the radio remnant is higher than that
expected from a SNR in the Sedov (1959) phase of evolution expanding
into a uniform, isotropic interstellar medium ($R\propto t^{2/5}$). On
the other hand, the observed value is lower than the expansion
parameter expected for a Sedov-phase SNR expanding into a steady
stellar wind ($\rho \sim r^{-2}$) density profile ($R\propto
t^{2/3}$).  However, neither of these models is strictly applicable,
since Kepler's SNR is believed to be in an earlier evolutionary phase
than Sedov, where, after all, the ejecta are supposed to be
dynamically unimportant.

The X-ray emission of Kepler's SNR is dominated by emission lines from
Si, S, Ar, Ca, and Fe (Becker et al.~1980; Kinugasa \& Tsunemi 1994)
with inferred abundances that are highly enriched compared to the
solar values (Hughes \& Helfand 1985). The conclusion that a
significant fraction of the X-ray emission comes from shock-heated SN
ejecta is compelling (Decourchelle et al.~1997). The X-ray--emitting
material constitutes far more mass than that visible in the optical or
radio band and so studies of the dynamics in the X-ray band are
particularly important for understanding the evolution of the remnant.
High angular resolution X-ray imaging is currently available only in
the soft X-ray band (photon energies below $\sim$2 keV), where the
flux from Kepler is overwhelming dominated by Fe L-shell ($n$p -- 2s
transitions) and Si XIII K$\alpha$ line emission.

X-ray expansion measurements of young SNRs are now possible for a
couple of reasons. First, the \rosat\ high resolution imager (HRI)
operated for long enough that significant time baselines between
individual \rosat\ observations of a few remnants are
available. Second, the HRIs on \rosat\ and \einstein\ are similar
enough that it is practical to take advantage of the considerably
longer time baselines between pairs of observations taken by these
different observatories.  The Kepler radio expansion rates suggest an
increase in the size of the X-ray remnant of $\sim$$0\myarcsec5$ over
a 5 year time period, which is close to the limit of what is possible
to detect with available HRI data.  For Kepler's SNR good data from
both X-ray observatories are available and in particular deep first
and second epoch observations separated by 5.5 yr were made by \rosat.
As I show below, agreement of the results determined from the various
combinations of \rosat\ and \einstein\ data provides a considerably
higher level of confidence in the results than any individual pair of
observations do.

\section{Observations}
\subsection{Initial Data Reduction}

Table 1 provides a log of all the high resolution \einstein\ and
\rosat\ imaging observations of Kepler's SNR.  The columns list the
observatory, start date, the Modified Julian Day (MJD) corresponding
to the average date of the observation, and the effective duration.

The remnant was observed twice by the \einstein\ high resolution
imager (EHRI): first on 1979 Sep 29 and then later on 1981 Mar 22 for
live-time corrected exposures of 5593.9 s and 12064.0 s,
respectively. In the following I refer to these observations as E1 and
E2.  The background level in each observation was estimated separately
and a value of $5.7\times 10^{-3}$ cts s$^{-1}$ arcmin$^{-2}$ was
found to be consistent with both.  There were roughly 8500 total
detected events above background from Kepler's SNR in observation E1
and 16,400 in E2 for rates of 1.51 s$^{-1}$ (E1) and 1.36 s$^{-1}$
(E2).  The difference in rates is believed to be due to a reduction in
the sensitivity of the EHRI during the course of the \einstein\
mission (Seward 1990).

Kepler's SNR was first observed by the \rosat\ HRI (RHRI) beginning on
1991 Sept 11 for a live-time corrected exposure of 36143.2 s and the
observation was completed about a day later. I refer to this as R1 in
the following. Due to the higher sensitivity and better calibration of
the RHRI compared to the EHRI, plus the much deeper
\rosat\ exposures resulting in more than 10 times as many detected
events, I have carried out a more detailed reduction of the \rosat\
data.  Throughout the entire data reduction process the pulse height
range of the R1 data was restricted to include only channels 1 to 9.
This reduced the background level significantly (by roughly 12\%),
while it hardly changed the number of X-ray events from the source
(reduced by 1.2\%). Aspect drift during the observation was a concern
that was addressed in the following manner. A separate image was made
from each of 15 time intervals (each typically 2000 s long)
corresponding to individual orbits during the observation.  Since
there were no bright X-ray point sources in the field, the remnant
itself was used to align the individual images using the IRAF task
crosscor (in package stsdas.analysis.fourier).  The initial
registration of the individual maps from the standard analysis was
fairly good: all of the individual maps were already aligned to within
1$^{\prime\prime}$ or better.  The images from all the sub-intervals
were registered to the nearest $0\myarcsec5$ pixel, shifted, and
added.

The second epoch RHRI observation (R2 hereafter) began on 1997 Mar 10,
ended on Mar 18 and included 67895.3 s of live-time corrected
exposure.  For this observation I used a range of pulse height
channels from 1 to 6,\footnote{The smaller range of pulse height
channels used here is due to an erroneous setting of the HRI high 
voltage, which applied to data taken between March 3 and May 6, 1997.
The effect of this was to shift the mean of the pulse height
distribution to smaller values by $\sim$2 channels. Based on
in-flight calibration measurements, this had no apparent change on the
quantum efficiency of the detector.} 
which reduced the background level by 16\% and
the number of source X-ray events by 1.4\% compared to the entire
range of pulse height channels. 
Here I divided the data into 36
individual time intervals to check and then compensate for any aspect
reconstruction errors.  Most of the individual maps were already
aligned to within better than 1$^{\prime\prime}$, but there were 3
separate intervals which produced offsets in declination of
2$^{\prime\prime}$--3$^{\prime\prime}$.  Again the images were
registered to the nearest $0\myarcsec5$ pixel, shifted, and added. For
both epochs, this shift-and-add alignment technique produced images
with a noticeably improved point response function.

Since the nominal RHRI background level integrated over the area of
the SNR represents slightly more than 1\% of the remnant's count rate,
I exerted some effort to estimate accurate values of the background
level in the two pointings. This estimate is made difficult by the
significant grain scattering halo from Kepler's SNR (Mauche \&
Gorenstein 1986; Predehl \& Schmitt 1995) and in fact careful
inspection of the data reveals that the entire RHRI field of view
contains some emission from the scattering halo. (Because of the
higher background the scattering halo is much less prominent in the
\einstein\ observations.)  To estimate the RHRI background level I
fitted a spatial power-law component plus a constant background level
to the surface brightness profile over the $2\myarcmin.5$ to 15$^\prime$
radial range. The fitted power-law components were consistent between
the two pointings (index of $-$1.5), although the background levels
differed by some 40\%, $3.2\times 10^{-3}$ cts s$^{-1}$ arcmin$^{-2}$
for the first pointing and $2.4\times 10^{-3}$ cts s$^{-1}$
arcmin$^{-2}$ for the second pointing.  This difference is within the
variation observed from field to field for the RHRI (David et
al.~1998).

The two RHRI observations were separated in time by roughly a
half-integer number of years, so the roll angles of the two pointings
differ by nearly 180$^\circ$. This means that spatial variations in
the efficiency of the detector will not cancel out between the two
pointings, so it was necessary to make an exposure map for each
observation.  The table of pointing directions as a function of time
(supplied as part of the \rosat\ standard processing) was binned into
a map containing the fractional amount of time the telescope spent
viewing differing positions on the sky during each observation.  This
produced a roughly rectangular region of nearly uniformly sampled sky,
4$^\prime$ long (in the direction of the wobble) and about 1$^\prime$
wide. The average roll angles of the pointings were determined:
$-$356$^\circ$ (R1) and $-$173$^\circ$ (R2).  A well-sampled image of
the bright earth was used to account for the spatial variation of the
quantum efficiency of the HRI, including the effects of an imperfect
gapmap.  The bright earth map was rotated about the optical axis and
then convolved with the sky viewing map to produce the exposure
map. Linear interpolation was used to sample the map onto different
pixel grids as appropriate.  Over the portion of the field containing
the image of Kepler's SNR the ratio of exposure between the first and
second epochs varied between 0.975 and 1.02.

The exposure- and deadtime-corrected, background-subtracted RHRI count
rates of Kepler's SNR are $5.491\pm 0.012\,\rm s^{-1}$ and $5.433\pm
0.009\,\rm s^{-1}$ from the first and second epoch images,
respectively, within a radius of 4$^\prime$.  Because of the remnant's
significant interstellar grain scattering halo, care was taken to make
the regions used for extracting the source counts as similar as
possible between the two epochs.  From the measured count rates it
would therefore appear that the total X-ray flux of Kepler's SNR has
declined by $\sim$1\% over the course of 5.5 yrs.  This, however, does
not account for any possible changes in the instrument's effective
area over the same period. Temporal variations in the effective area
of the RHRI have been monitored over the course of the mission by
regular observations of the SNR N132D in the Large Magellanic Cloud,
which is not expected to be variable on short ($\sim$10 yr)
timescales. The available data (David et al.~1998) do not show any
obvious secular trend, although the scatter in the observed count
rates is considerably larger (RMS scatter of $\sim$ $\pm3$\%) than the
statistical uncertainty in the individual measurements themselves
(typically $\pm1$\%).  This indicates that some other effects may be
influencing the RHRI's overall effective area, perhaps on an
observation-to-observation basis.  I conclude, therefore, that
Kepler's SNR shows no significant evidence for a change in X-ray flux
although changes at the level of a few percent are possible.

\subsection{Registration}

Five weak X-ray point sources appeared in both the R1 and R2 images
between 4$^\prime$ and 12$^\prime$ off-axis. I used these sources to
register the different \rosat\ images to each other (including
position shifts as well as a rotation) and to constrain any changes to
the RHRI plate scale empirically.  According to the \rosat\ standard
processing there are no SIMBAD sources within 20$^{\prime\prime}$ of
the nominal positions of these sources.  I have not pursued the
determination of absolute positions or the identification of optical
or radio counterparts, since this information is not necessary for
image registration.

The positions of the five sources in both observations were determined
by fits of the appropriate off-axis RHRI point response function
(David et al.~1998) to the image data, using a likelihood function
derived for Poisson-distributed data as the figure-of-merit
function. Pixel positions as well as their statistical uncertainties
were determined.  The differences between the fitted positions in the
two epochs were compared using $\chi^2$. A single positional offset is
a good fit to the five point sources: the $\chi^2$ associated with
this fit is 8.1 for 8 degrees of freedom.  The best fit difference in
the pixel positions between the two epochs are $5\myarcsec0 \pm
0\myarcsec4$ (in right ascension) and $-7\myarcsec6 \pm 0\myarcsec3$
(in declination).  The sense of the difference is that the second
epoch image is south and east of the first.  Figure 1 plots the
individual offsets and uncertainties from the five point sources.  The
filled dot symbol shows the best fit positional offset and the dashed
curve is the 1 sigma error contour (at $\Delta\chi^2 = 2.30$ as
appropriate for two interesting parameters).  There is no evidence for
a relative rotation between the two epochs.  If a rotation about the
optical axis is allowed in addition to the positional shifts, the best
fit rotation angle is found to be $0\mydegree.04$ and the $\chi^2$
reduces to 7.5, which is not a statistically significant reduction
relative to the unrotated fits.

\begin{figure*}
\plotfiddle{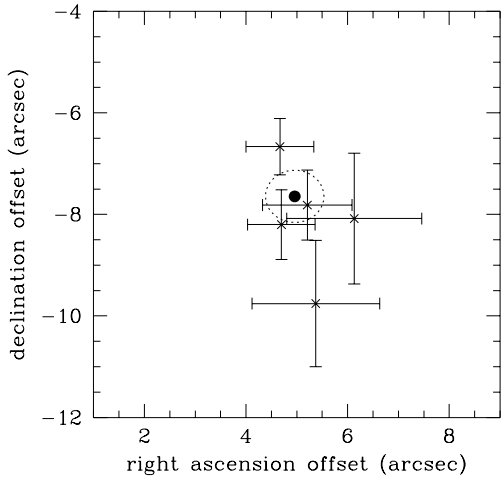}{2.5in}{0}{50}{50}{-150}{-125}
\figcaption[fig1.ps]{Positional offsets and uncertainties for five
X-ray point sources serendipitously located in the field of view
of both the first and second epoch HRI data of Kepler's SNR. The best
fit position for no relative rotation between epochs is shown as the
filled dot symbol, while the dashed contour shows the 1 sigma uncertainty
on that position.
\label{fig1}}
\end{figure*}

\begin{figure*}[b]
\plotfiddle{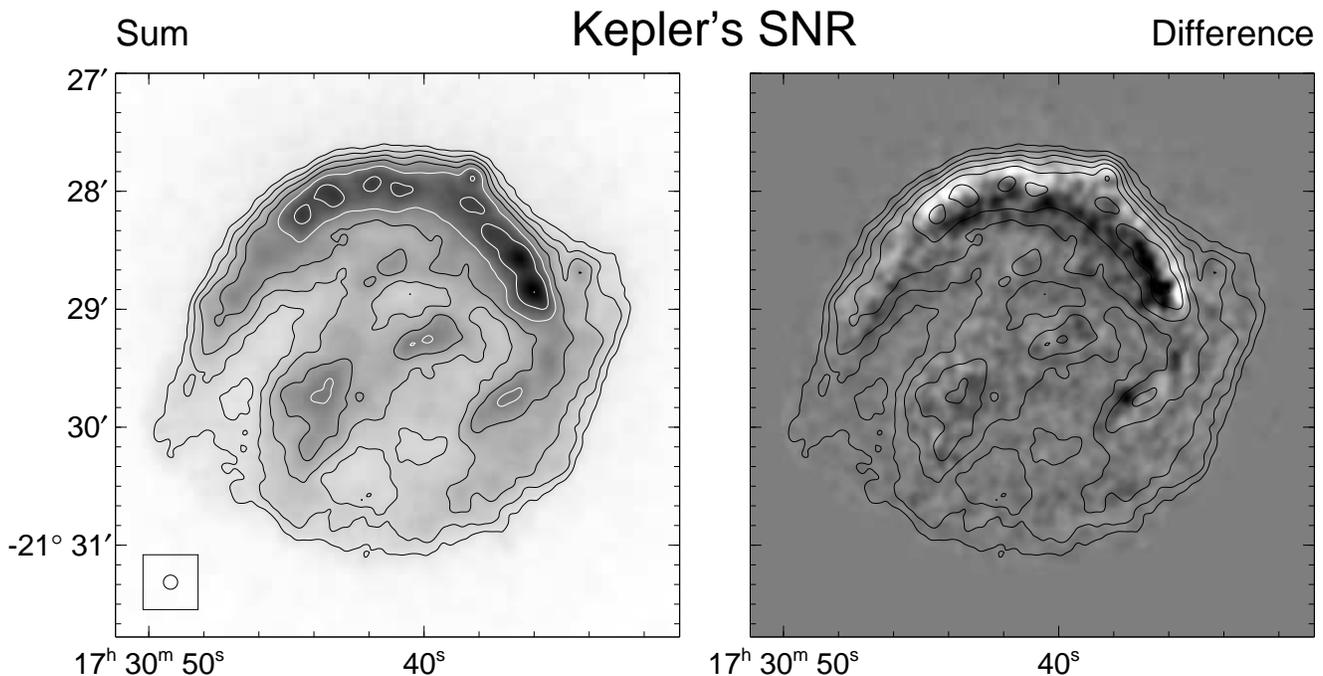}{3.25in}{90}{75}{75}{300}{-90}
\figcaption[fig2.ps]{\rosat\ HRI image of Kepler's SNR.  The left
panel shows the summed data from two observations with a total
exposure of 104 ks. The data were adaptively smoothed to an
approximate signal-to-noise ratio per pixel of 5.  This grayscale
image here uses a square-root scaling for the image display.  The
contour values are logarithmically spaced by multiplicative factors of
1.864 and start at a value of 0.0865 counts s$^{-1}$
arcmin$^{-2}$. The effective resolution of the map, including both the
instrumental PSF and the average width of the smoothing kernel,
is shown at the lower left. The right panel shows the difference
between the two observations, which were separated in time by 5.5 yr.
Before subtraction the separate images were registered to the nearest
$0\myarcsec5$ pixel and scaled by their respective exposure times.
The grayscale image on this part of the figure uses a linear scaling
for the image display and values are displayed between $\pm$0.36 counts
s$^{-1}$ arcmin$^{-2}$.  Dark regions are places where the count rate
has dropped between the first and second epoch images and the light
colored regions are places where the rate has increased. The contours
of total X-ray intensity, which are the same as on the left panel, are
shown for reference.
\label{fig2}}
\end{figure*}

These sources also provide a limit on any changes to the RHRI plate
scale over the 5.5 years that separate the first and second epoch
images. I find that decreasing the plate scale (i.e., the number of
arcseconds per pixel) of the second epoch observation by $0.16\% \pm
0.10\%$ (1-sigma uncertainty) resulted in a slightly better fit
($\chi^2 = 5.3$) to the relative positions of the point sources in the
field.
Using the positions of X-ray sources associated with globular clusters
in M31 F.~Primini (1995, private communication) has shown that the
plate scale in July 1990 was $0.4992 \pm 0.0004$
$^{\prime\prime}$/pixel, while in July 1994 it was $0.4993 \pm 0.0002$
$^{\prime\prime}$/pixel which represents an {\it increase} of $0.02\%
\pm 0.09\%$ in the \rosat\ HRI plate scale over the course of 4 years.
My intent here is not to claim that there has been a significant
change in the plate scale of the \rosat\ HRI (indeed my result is
significant at only about the 90\% confidence level), but rather to
show that even this weak limit is still about a factor of 10 less than
the expansion I measure for Kepler's SNR. Below I use Primini's and my
values for the change in the RHRI plate scale as one estimate for the
systematic errors on the expansion measurement.

\begin{deluxetable}{ccc}
\tablecaption{Global Mean Percentage Expansion of Kepler's SNR}
\tablewidth{4truein}
\tablehead{
\colhead{Datasets} & \colhead{Time Difference (yr)} &
\colhead{Expansion (\%)}
}
\startdata
R1--R2  &  $\phantom{0}5.51$ & $1.39^{+0.12}_{-0.13} \,^{+0.07}_{-0.18}$\nl
E2--R1  &  $10.47$  & $2.21^{+0.25}_{-0.26} \,^{+0.18}_{-0.05}$\nl
E1--R1  &  $11.95$  & $2.88^{+0.44}_{-0.33} \,^{+0.34}_{-0.11}$\nl
E2--R2  &  $15.98$  & $3.68^{+0.22}_{-0.23} \,^{+0.11}_{-0.20}$\nl
E1--R2  &  $17.46$  & $4.49^{+0.30}_{-0.33} \,^{+0.16}_{-0.18}$\nl
\enddata
\end{deluxetable}

The five point sources allow accurate relative registration of the two
\rosat\ pointings. In figure 2, I show the sum of the two RHRI images
as well as their difference after being registered and scaled by the
ratio of live times. The latter shows a rim of positive emission
around most of the periphery of the remnant with a region of negative
flux further in, a clear signature of expansion.

\section{Dynamics of the X-Ray Remnant}
\subsection{Global Mean Expansion Rate}

The four \einstein\ and \rosat\ observations give six different time
baselines varying from $\sim$1.5 yrs to nearly 17.5 yrs over which it
is possible to measure the expansion of Kepler's supernova remnant.  I
use only the five longer baselines, since the shortest one, which
corresponds to the two \einstein\ pointings, is also the one with the
smallest number of detected events and thus is the least
sensitive. Table 2 lists the five pairs of measurements used and the
difference in time between the average date of each observation.

There are a number of complications associated with the comparison of
these pairs of X-ray measurements.  First, except for the two \rosat\
pointings, there are no serendipitous point sources in the field that
can be used for relative position registration. Second, small-scale
spatial features in the remnant are neither bright enough nor distinct
enough to be directly tracked from one observation to another. However
Figure 2 clearly shows that there has been a significant change in the
remnant during the time interval between the two \rosat\ pointings.
In the following I investigate whether these changes are consistent
with the remnant undergoing expansion and determine numerical values
for the rate of expansion.

The approach I take is to use the image from one of the observations
as the ``model'' for the X-ray emission from the remnant, which is
scaled in intensity, shifted in position, and expanded or contracted
in spatial scale to match the image from another observation, referred
to as the ``data.'' Note that for this part of the study there is no
need to choose an expansion center.  The model assumes that the
expansion rate is uniform over the entire image of the remnant, both
radially and azimuthally, and thus what I derive is a global mean
expansion rate. Furthermore, this method is not sensitive to an
overall translation or proper motion of the entire remnant. Any such
translation is inextricably tied up with the relative registration of
the various images (which in general is known to an accuracy of no
better than 5$^{\prime\prime}$-10$^{\prime\prime}$ based on absolute
positions from aspect reconstruction) and thus cannot be determined
separately (for the moment I am ignoring the accurate registration
between the two RHRI images, however see \S 3.2 below).  Although the
choice of which observation to use as the model or the data is
arbitrary, in practice the deeper, later epoch observation is taken to
be the model image. The figure-of-merit function for the model-data
comparison is a maximum likelihood estimator derived for
Poisson-distributed data. This is necessary since even for the second
epoch \rosat\ observation, which contains the largest number of
detected X-rays, the most probable number of events in any
1$^{\prime\prime}$ square pixel over the central 2$^\prime$ image of
Kepler's SNR is 0, while the average number of events per pixel is
$\sim$8. As is well known, this maximum likelihood figure-of-merit
function does not provide an analytic goodness-of-fit criterion.
However the function was implemented in such a way that {\it
differences} in best-fit values were distributed like $\chi^2$,
through the ``likelihood ratio test'' (see, for example, Kendall \&
Stuart 1979), which provided the basis for estimating statistical
uncertainties. This technique of model fitting to two-dimensional
imaging data and the software developed for it has been used in a
number of situations; see Hughes \& Birkinshaw (1998) for another
example of its application and a more detailed description. The new
software for this project was verified in a number of ways as
explained below.

Some processing of the model image was unavoidable.  In order to shift
the image by a fractional number of pixels and expand or contract the
spatial scale, some type of interpolation over the image had to be
done. I used bilinear interpolation in right ascension and declination
on a relatively smooth version of the model image.  For the comparison
of the two \rosat\ pointings, the model image was adaptively smoothed.
This was done by splitting the model data into a number of separate
images according to surface brightness and smoothing each separate
image with a different gaussian smoothing scale.  The smoothing scales
were chosen so that roughly the same number of counts on average were
contained within a smoothing scale length in all the separate images.
It is essential that the smoothing process not introduce any bias into
the expansion measurement. This was verified by taking the same
observation (R2) as both the data and model, carrying out this
smoothing process on the model, and fitting for the relative position
shift, the change in spatial scale, and the intensity change.  The
best fit relative position shift was $<$0.001$^{\prime\prime}$ and the
best fit changes in spatial scale and intensity were both $<$0.01\%,
clearly indicating that the smoothing process has not introduced a
bias into the fitted results.  

I also rigged up a test to verify that this technique could accurately
detect changes in the sizes of X-ray images (as opposed to a null test
like above). The smoothed version of observation R2, which was
constructed with $1^{\prime\prime}$ square pixels, served as the
model. For the data the same observation, but blocked to $1\myarcsec5$
square pixels, was used.  The software carries out the fits in terms
of image pixels, so the two images clearly differ significantly in
spatial scale and relative normalization.  Nevertheless the software
returned precise and accurate best-fit values for the spatial scale:
$0.6664\pm 0.0002$ (expected value: 2/3) and relative normalization:
$2.2473\pm 0.0014$ (expected value: 2.25) (1-sigma errors).  This test
and the others described in this section give confidence in the
software and technique used for these expansion measurements.

For the comparison of the \einstein\ and \rosat\ data, the most
critical image processing step was to account for the differing
on-axis point response functions (PRFs) of the two observatories. The
\einstein\ PRF is somewhat broader and has a stronger scattering halo
than the \rosat\ one. In addition the \einstein\ PRF has some
dependence on photon energy, while the \rosat\ one is nearly
achromatic.  Parametric models of the PRFs for both instruments are
available in the literature (\einstein: Henry \& Henriksen 1986;
\rosat: David et al.~1998).  As the first step, the Lucy-Richardson
(Richardson 1972; Lucy 1974) algorithm was used to deconvolve the
\rosat\ PRF from the RHRI data.  Beyond a certain point the results
were relatively insensitive to the number of iterations of the
algorithm; five or ten iterations were found to produce nearly
identical results.  The deconvolved image was then convolved with the
\einstein\ PRF to produce a model image with a spatial resolution
approximating that of the EHRI. This process produced a smooth enough
model that no additional smoothing step was necessary. In general the
fits of the model image were quite good (after accounting for the
expansion of the remnant) and it is clear, at least qualitatively,
that there was no gross mismatch in the PRFs.  Toward the end of this
section (see discussion of SNR N132D below) I describe how I verified
quantitatively that the method just described does not bias the
expansion measurement.

This now brings me to the discussion of the fits themselves. The first
set of fits was done under the assumption that the remnant had not
undergone any expansion.  There were four parameters in these fits:
relative position shift in right ascension and declination, ratio of
intensity (or normalization), and background level. The data were fit
over a circular portion of the HRI field with a radius of
$2\myarcmin.5$ centered on the remnant. The results are summarized in
the set of plots at the top of figure 3 which show the
azimuthally-averaged surface brightness profiles for the five pairs of
observations considered.  Although the gross match between the data
and model is good, plots of the residuals (second epoch minus first
epoch) show an ``S''-shaped pattern near the remnant's edge that is a
characteristic signature of expansion.

\begin{figure*}
\plotfiddle{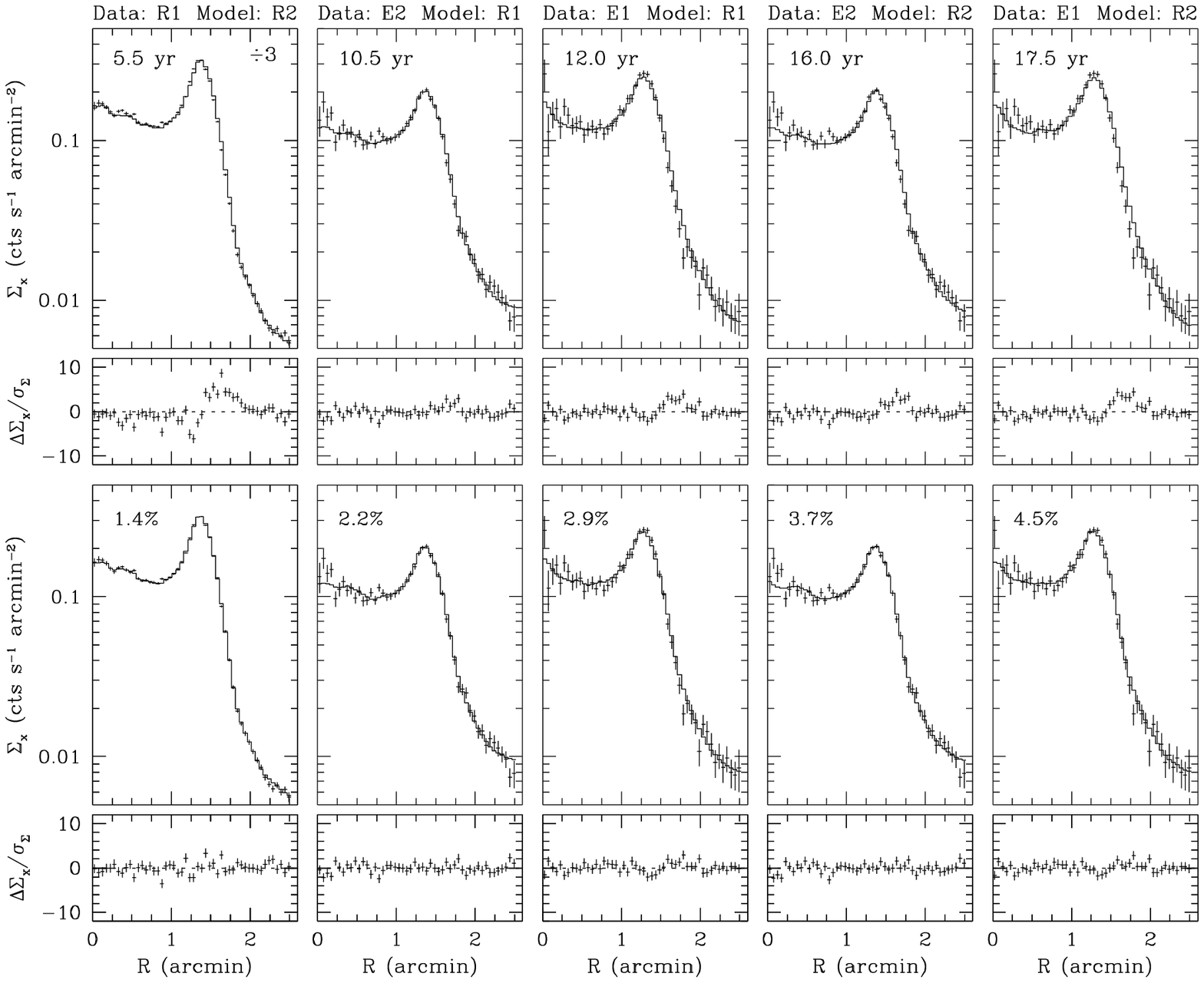}{4.1in}{0}{70}{70}{-240}{-120}
\figcaption[fig3.ps]{Azimuthally-averaged radial X-ray surface
brightness profiles of Kepler's supernova remnant showing a pairwise
comparison of observations taken at different times. The $\Sigma_X$
panels plot one observation as the ``Data'' (points with error bars)
while another observation, after being shifted in position and scaled
in intensity to match, is plotted as the ``Model'' (histogram). The
various combinations of data and model are labelled at the top; note
that the observation taken to be the ``Data'' is always the earlier
one. The $\Delta\Sigma_X/\sigma_\Sigma$ panels show the residuals
between the model and data. The set of panels at the top assume that
the remnant has undergone no expansion over the period of time between
observations (values as indicated).  Note the characteristic pattern
of residuals indicative of expansion of the remnant.  The bottom set
of panels assume that the remnant has expanded between observations
and the best-fit expansion values for each pair of measurements are
shown. In all cases these fits are a considerable improvement over the
fits assuming no expansion.  The \rosat\ surface brightness data
(first pair of panels on the left) have been divided by a factor of 3
in order to fit on the same scale at the \einstein\ data (remaining
panels).
\label{fig3}}
\plotfiddle{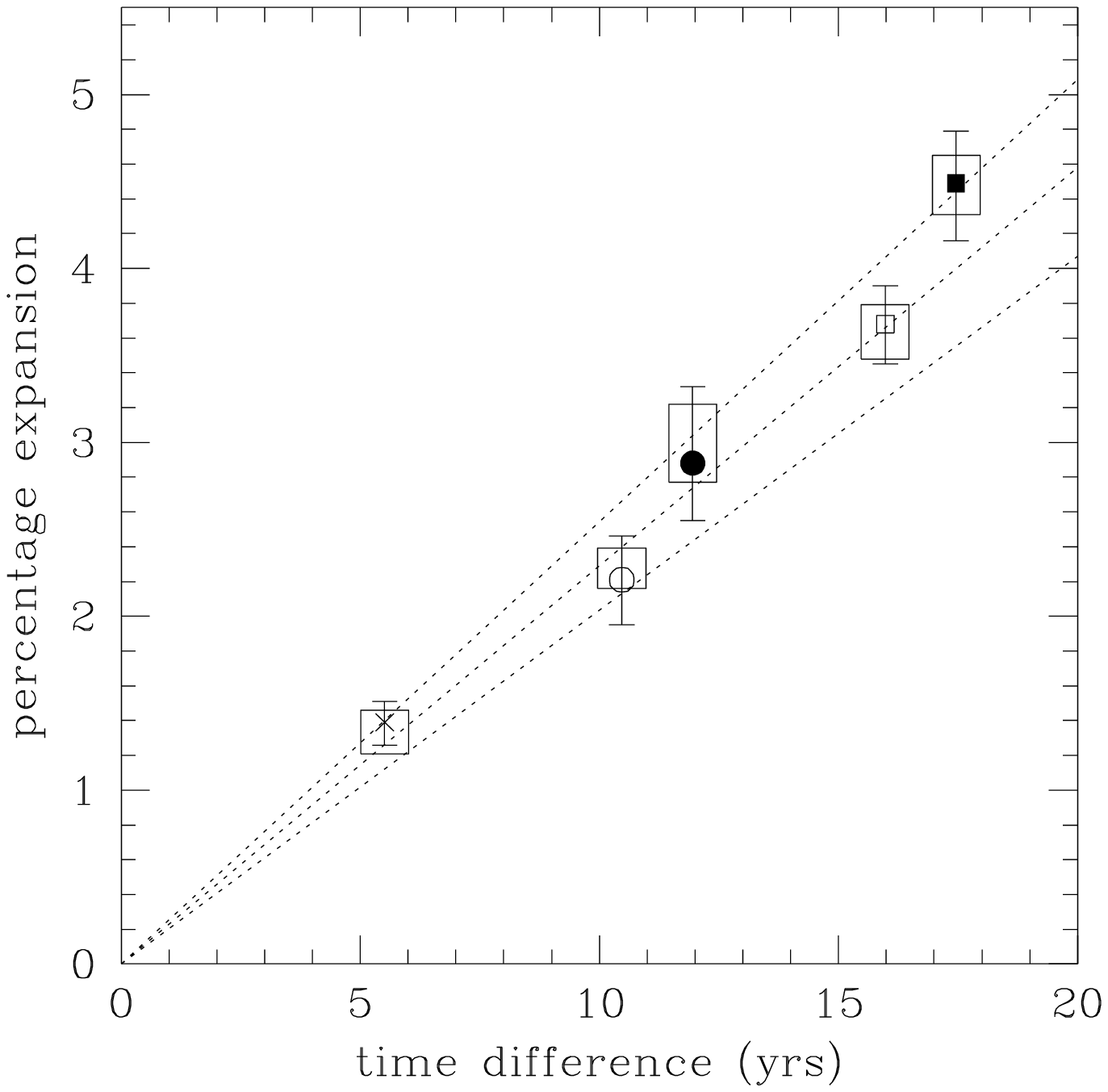}{2.9in}{0}{50}{50}{-175}{-100}
\figcaption[fig4.ps]{Global mean percentage expansion of Kepler's
supernova remnant versus the time difference between imaging
measurements. The five points come from using different pairs of
measurements. The cross symbol, which indicates the shortest time
difference, comes from comparing the two {\it ROSAT} pointings. The
circles compare the {\it ROSAT} first epoch pointing with the {\it
Einstein} second epoch (open symbol) and first epoch (filled symbol)
observations. The squares show the expansion rate using the {\it
ROSAT} second epoch pointing with the {\it Einstein} second epoch
(open symbol) and first epoch (filled symbol) observation.  The error
bars show the statistical uncertainty, while the boxes that surround
each data point give an estimate of the systematic uncertainty.  The
dotted curves show the expansion rate expected assuming expansion
parameters of 0.8 (bottom curve), 0.9 (middle) and 1 (top).
Remarkably, the X-ray remnant of Kepler's SNR appears to be expanding
at close to the free expansion rate.
\label{fig4}}
\end{figure*}

In the next set of fits the mean expansion of the remnant expressed as
a percentage was included as one additional free parameter.  The
resulting surface brightness profiles are shown in the plots at the
bottom of figure 3.  In all cases the quality of the fit improved
significantly with the residuals now appearing quite flat.  The
reduction in the value of the likelihood function was 56 for the least
significant case (E2 compared to R1) while it was over 800 for the
most significant one (R1 compared to R2).  According to the likelihood
ratio test, as mentioned above, the distribution function for the
change in the likelihood function in this case is given by a $\chi^2$
distribution with a single degree of freedom. Consequently the
confidence levels associated with these reductions in the likelihood
function are all off-scale (i.e., $\gg$ 3 sigma).

The best fit percentage expansion factors are plotted versus the time
difference between observations in figure 4 and numerical values are
presented in column 3 of table 2. The uncertainties quoted in the
table consist of 1-sigma statistical uncertainties (the first pair of
values) as well as systematic uncertainties (the second pair of
values). The error bars in figure 4 show the statistical
uncertainties, while the boxes drawn about each symbol indicate the
level of systematic uncertainty.  The five individual values are
consistent at about the 1 sigma confidence level, assuming only
statistical uncertainty, with an ensemble expansion rate of
$\sim$0.24\% yr$^{-1}$.  However, these five individual values are not
independent, so they cannot be combined directly.  I can construct an
ensemble average expansion value from independent pairs of
observations in two different ways: E2-R2 and E1-R1, or E1-R2 and
E2-R1. As a conservative estimate I take the maximum absolute value of
the systematic uncertainty within each pair as the systematic
uncertainty to associate with the ensemble value.  These result in
expansion rates of $0.239^{+0.015\,+0.017}_{-0.015\,-0.010}\,\%\, \rm
yr^{-1}$ (E2-R2 and E1-R1) and
$0.232^{+0.013\,+0.028}_{-0.013\,-0.013}\,\%\, \rm yr^{-1}$ (E1-R2 and
E2-R1), where the first set of quoted errors comes from statistical
uncertainty and the second set incorporates systematic effects.
Because the former value has a smaller systematic uncertainty range
(and since the two ensemble averages are fully consistent with each
other at the 1-sigma level), I will consider the former value to
represent the global mean expansion rate of Kepler's SNR in the
remainder of this article.

The quoted statistical errors are based on three sources of random
noise: Poisson noise in the data image, Poisson noise in the model
image, and the statistical uncertainty on the determination of the
relative plate scale between pairs of observations.  The total
statistical error I quote is the root-sum-square combination of the
three components.  The first term was found in the usual manner by
searching for values of the expansion that correspond to an increase
in the figure-of-merit function of +1 above the best fit case (1 sigma
confidence level).  The second error term was assessed using a Monte
Carlo approach which was carried out for each pair of data-model
images.  Simulated data were created from the adaptively smoothed
\rosat\ image that corresponds to the model using a Poisson random
number generator, assuming the same exposure time as in the original
image.  For comparison with the \einstein\ data this simulated image
was then run through the deconvolution and convolution steps discussed
above that account for the differing instrumental PRFs.  For
comparison with the other \rosat\ data the simulated image was
adaptively smoothed.  The best fit expansion value was determined
using the processed, simulated image as the model and then the whole
process was repeated until a total of 10 separate random realizations
of the model were completed. The statistical error associated with the
model was taken as the root-mean-square of the distribution of these
best fit expansion values.  In general for the \rosat-\einstein\
comparison this term was a factor of 2 to 3 times smaller than the
error from the Poisson noise in the data itself and thus does not
cause a significant increase in the total error.  However for the
\rosat-\rosat\ comparison this term was comparable to the error from
the noise in the data itself.  The final source of statistical error
was from the uncertainty in measuring any change in the plate scale
between observations.  For the \rosat-\rosat\ comparison the value of
$\pm$0.10\% uncertainty derived in \S 2.2 results in an identical
uncertainty on the expansion of the remnant.  This turns out to be the
dominant source of random error for this pair of observations. I use a
value of $\pm$0.07\% for the error in determination of the relative
plate scale between \rosat\ and \einstein\ (David et al.\ 1998), which
results in an error of the same size for the expansion.  This error is
small in comparison to the Poisson noise in the \einstein\ data.

Systematic uncertainties have the potential to bias the expansion
measurement, so some effort was made to investigate them.  For all
pairs of data I studied the effects of varying the spatial region of
the image over which the fits were done.  The center of the circular
fit region was shifted by 10$^{\prime\prime}$ in each of the four
cardinal directions and the radius of the fit region was changed by
$\pm$$0\myarcmin.5$.  A new best-fit expansion value was obtained for
each of these cases separately and the systematic uncertainty was
taken to be the difference between the new value and the nominal
best-fit value.  The uncertainty due to the center shift was small,
less than 0.06\%, while that due to the size of the fit region ranged
from 0 to a value as large as 0.2\%.  For the \rosat-\einstein\
comparison two other effects were investigated: the number of
iterations of the Lucy-Richardson deconvolution algorithm (between 5
and 10, which resulted in an uncertainty range of 0--0.07\%) and
changes in the modeled EHRI PRF.  The functional form that describes
the EHRI PRF depends on three parameters: two spatial scales and a
relative normalization. Based on ground calibration data, the values
of these parameters vary by roughly 10\% as a function of photon
energy over the $\sim$0.5--2 keV band and so this is the level of
uncertainty I assume. Each of the three parameters was separately
varied from its nominal value by $\pm$10\% and new best-fit
expansion values were obtained. The systematic uncertainties from this
effect were all less than 0.07\%. Finally for the \rosat-\rosat\
comparison, the change in plate scale was included as an
uncertainty. According to the discussion in \S 2.2, the systematic
errors resulting from possible changes in the RHRI plate scale are
+0.02\% and $-$0.16\%, which are the dominant uncertainties for this
pair of images.  The several systematic errors are combined by direct
summation to produce the values shown in table 2 and figure 4.

An additional possible source of systematic uncertainty arises from
the difference in the effective area functions of \einstein\ and
\rosat.  Because of the relatively high absorbing column density to
Kepler's SNR (estimates in the literature range from $\sim$$3\times
10^{21}$ cm$^{-2}$ to $\sim$$6\times 10^{21}$ cm$^{-2}$) the X-ray
band below roughly 0.8 keV is largely absorbed away, so that the main
difference in the effective area functions is \einstein's greater
sensitivity to X-rays with energies above 1.8 keV (see Vancura,
Gorenstein, \& Hughes 1995).  Thus spatial variations in the X-ray
spectral emissivity of the remnant could produce apparent differences
in the relative flux seen by the EHRI compared to the RHRI. A strong
radial gradient in the effective hardness of the remnant would be
particularly pernicious since it would tend to appear predominantly as
a radial increase or decrease in flux when the EHRI and RHRI data were
compared that might masquerade as an expansion or contraction of the
remnant. Although clearly a source of systematic uncertainty, it is
unlikely that this effect is a significant source of bias in the
results.  The strongest arguments against this effect being important
are (1) the good linear relationship found between the derived amount
of expansion and the time baseline between observations (figure 4) and
(2) the excellent agreement between the expansion rate for the
\rosat-\rosat\ comparison (where there is no spectral band difference)
and the \rosat-\einstein\ comparison.

As an additional check on these overall results I carried out the same
analysis for the SNR N132D in the Large Magellanic Cloud.  Both the
\einstein\ and \rosat\ HRIs observed this remnant with the RHRI
pointing being about a factor of 10 deeper than the EHRI one. The
observations were separated by a span of roughly 11.6 yrs.  N132D is a
middle-aged remnant, about 4000 yrs old, roughly 12 pc in radius, with
a shock velocity of approximately 800 km s$^{-1}$ (Hughes, Hayashi, \&
Koyama 1998).  Over a span of 11.6 yrs, it should have expanded by a
negligible amount, less than 0.1\%.  Following the same procedures and
software as used for Kepler's SNR I get an expansion rate for N132D of
$-1.1^{+0.29}_{-0.25}\,^{+0.44}_{-0.55}$\%, where again the first set
of errors denote the statistical uncertainty and the second set denote
the systematic uncertainty. The sign of the expansion indicates that
the second epoch image of N132D (the RHRI one) is slightly smaller
than the first, which is opposite to what was found for Kepler's SNR,
as well as somewhat implausible.  The result may be due to the effect
of spatial variations of the X-ray spectral emissivity from the
remnant coupled with the different bandpasses of the \einstein\ and
\rosat\ HRIs.  However, when the systematic errors are considered, the
measured value of expansion for N132D is not significant, lying  
2 sigma away from zero.  This study is a more severe test of the null
hypothesis for expansion because N132D is considerably smaller in
angular size than Kepler's SNR (45$^{\prime\prime}$ radius vs.\
100$^{\prime\prime}$ radius), so that systematic effects due to the
different point response functions are more significant.  Nevertheless,
the test confidently establishes that the expansion results on
Kepler's SNR are not an artifact of the analysis technique or
procedures.

\subsection{Expansion as a Function of Azimuth}

\begin{figure*}[b]
\plotfiddle{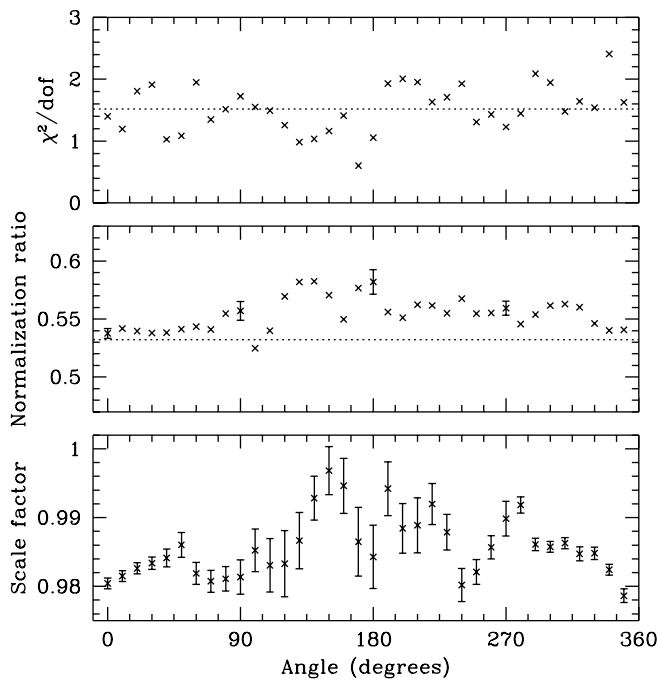}{3.5in}{0}{50}{50}{-150}{-70}
\figcaption[fig5.ps] {Variation of the best-fit spatial scale factor,
intensity normalization, and reduced $\chi^2$ with azimuthal angle
from a comparison of two \rosat\ HRI observations of Kepler's
supernova remnant separated by approximately 5.5 yrs.  Angles are
measured east from north and the error bars are 1 sigma and are purely
statistical.  The dashed curve in the top panel is plotted at the
reduced $\chi^2$ value corresponding to the 99\% confidence level. The
dashed curve in the middle panel corresponds to the ratio of detector
live times.
\label{fig5}}
\end{figure*}

The same software was used to determine the expansion rate as a
function of azimuthal position around the remnant. The two RHRI
observations were chosen for this part of the study since they were
the most sensitive pair and subject to the least number of systematic
errors. As before, the later epoch, adaptively-smoothed image served
as the model which was scaled in intensity and spatial scale to match
the earlier data.  Comparison between the datasets was done for
pie-shaped regions, centered on a fixed position (the expansion
center, see below) with each region having an outer radius of
$2\myarcmin.5$ and an angular extent of 20$^\circ$.  Separate fits
were done for 36 angular regions centered on angles of 0$^\circ$,
10$^\circ$, 20$^\circ$, etc.  The sign convention used in this paper
has angles increasing in a counterclockwise sense from north.

There were only two free parameters for each fit: the ratio of
intensity normalizations (which would equal the ratio of detector live
times if there were no changes in surface brightness) and the relative
spatial scale between the epochs. The background was kept fixed at the
nominal difference in background levels of $8\times 10^{-3}$ counts
s$^{-1}$ arcmin$^{-2}$. Unlike before, no position shift was included,
since the registration of the images is known, at least to
$\sim$$0\myarcsec5$ or so.  Instead the expansion is considered to be
strictly in the radial direction about a position in the remnant,
called the expansion center.  I chose to take the center of the X-ray
remnant at position 17:30:41 $-$21:29:23 (J2000) (determined by an
eyeball fit of a circle to the outer X-ray surface brightness
contours), as the location of the
expansion center.  In order to actually determine the expansion center
it would be necessary to track individual X-ray emission features and
determine the direction of their proper motion.  Extrapolating several
such tracks back to the center and noting where they intersect would
then provide an estimate of the position of the expansion center.
This is not yet possible to do in the X-ray band.

\begin{figure*}[t]
\plotfiddle{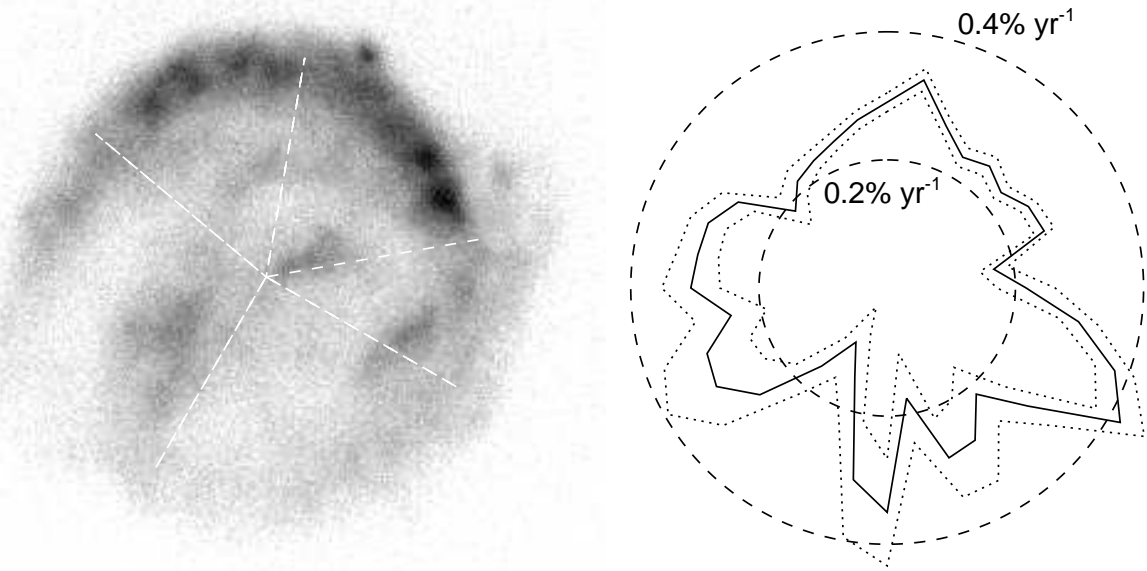}{3.6in}{90}{75}{75}{300}{-110}
\figcaption[fig6.ps] {The left side of the figure shows the X-ray
emission from Kepler's supernova remnant made by summing both \rosat\
images and displayed using a square root intensity scale. The image
shows the raw number of detected X-ray events per 1$^{\prime\prime}$
square pixel without background subtraction, smoothing, or exposure
correction.  The right side is a polar plot of the azimuthal expansion
rates (solid curve) and 1 sigma statistical error range (dotted
curves) from a comparison of the two \rosat\ HRI observations.  The
dashed circles indicate the scale of this part of the figure in units
of percentage expansion per year.  Several azimuthal directions that
correspond to features in the expansion rate plot are indicated on the
grayscale image of the remnant (see text).
\label{fig6}}
\end{figure*}

Figure 5 shows the results of these fits. The bottom panel shows the
fitted relative spatial scale, expressed as a decimal fraction,
required to make the second epoch data match the first epoch.  (The
scale factor is less than 1 because the second epoch image is larger
than the first.) The 1-sigma error bars shown are rather small
($\pm$0.08\%) in the bright northern part of the remnant but become
much larger ($\pm$0.5\%) in the faint southern part. They only include
the statistical uncertainty in the data.  The middle panel shows the
ratio of fitted normalizations, which with one exception are all
greater than the ratio of the detector live times (indicated on the
plot as the dotted line). This indicates that the remnant has
decreased in surface brightness between the two epochs and that the
fainter regions toward the south have decreased more in brightness
than the brighter northern regions.  However the brightness decrease
is offset by the increase in size of the remnant so that the total
count rate from the remnant is nearly the same for the two epochs.
Note that the (average) ratio of normalizations found here is
consistent with the value determined from the global fits discussed
above (\S3.1).  The top panel shows the $\chi^2$ per degree of freedom
(for approximately 50 total degrees of freedom) associated with each
best fit.  For this calculation only, both the data and scaled model
within each azimuthal sector were binned into a radial profile with 50
bins in radius.  These surface brightness profiles were all visually
inspected and no obvious systematic mismatch in the fits were
apparent.  However the reduced $\chi^2$ values shown in the figure
scatter about a mean value of $\sim$1.5, indicating that the fits in
general are formally unacceptable at about the 99\% confidence level.
It is likely therefore that other effects which have not been included
in this analysis, such as non-radial motions, flux changes on small
spatial scales, or radial variations in the expansion rate, are
important for Kepler's SNR.  The first two topics are beyond the scope
of this work, but the third topic is explored in the next section of
the paper.

In order to check the relative registration of the two images once
more, the fitted spatial scale factors, now expressed as a percentage
expansion, were themselves fitted using $\chi^2$ to a function
consisting of an azimuthally uniform expansion rate plus one that
varied with angle as a sinusoid. The inclusion of the second term was
highly statistically significant, resulting in a reduction in $\chi^2$
of at least 37 for the addition of two new free parameters. The best
fit was obtained for a uniform expansion rate of 1.4\% and a cosine
function with an amplitude of 0.4\% and phase angle of
207$^\circ$. The effect of the sinusoid was to decrease the expansion
rates in the north while increasing them in the south. The amplitude
of the sinusoid could be caused by a relative position shift of as
little as $0\myarcsec4$, which is of the same order as the computed
error in the relative registration.  Because of this, plus the
excellent agreement of the uniform expansion rate determined here with
the global mean value determined previously (\S 3.1), I have chosen to
remove the fitted sinusoid from the azimuthal expansion rates.  Note
that the sinusoidal term sets a limit of $\sim$$0\myarcsec07 \, \rm
yr^{-1}$ on the overall proper motion of the X-ray remnant,
corresponding to a limit on the transverse speed of $<$1660 $(D/5\,
\rm kpc)\, km\, s^{-1}$.  This should be interpreted only as a broad
guide to the level of proper motion of the remnant excluded by the
X-ray data since the amplitude of the sinusoid and thus the upper
limit are sensitive to the somewhat arbitrary choice of expansion
center made here.

The best fit model (uniform plus sinusoid) to the azimuthal expansion
rates is formally not acceptable yielding $\chi^2 = 51$ for 15 degrees
of freedom, indicating that there are significant azimuthal variations
on smaller angular scales.  This is obvious as well in figure 6 which
shows the azimuthal expansion rates (corrected for the sinusoidal
term) and the 1-sigma errors in a polar plot presentation. Next to it
is a grayscale depiction of the X-ray image of the remnant.  Some of
the prominent azimuthal variations in expansion rate are correlated
with emission features. The most rapidly expanding portion of the
remnant at a position angle of 240$^\circ$ contains a knot of emission
about two-thirds of the way out from the center.  This sector has an
apparent expansion rate of $0.42\pm0.04$\% yr$^{-1}$.  The two local
minima in expansion rates at 50$^\circ$ and 280$^\circ$ are correlated
with the edges of the bright northern portion of the shell.  Other
azimuthal variations do not appear to correlate with obvious emission
features.  These include the position of the peak expansion rate
($0.33\pm0.02$\% yr$^{-1}$) in the north at 350$^\circ$ and the
overall minimum in expansion rate ($0.10\pm0.06$\% yr$^{-1}$) at
150$^\circ$.  The pattern of azimuthal variation in expansion rates
shown in figure 6 appears to be robust against modest changes in the
position of the expansion center, as long as the sinusoidal term is
fitted and removed. A second set of fits made with the expansion
center displaced by about $2\myarcsec5$ gave results that agreed
within the 1-sigma statistical errors with the previous fits.
Finally, I point out that the overall scale of these azimuthal
expansion results is additionally uncertain by roughly
$^{+0.02}_{-0.05}$\% yr$^{-1}$ due to systematic errors of which the
plate scale uncertainty is the dominant one.

\subsection{Expansion as a Function of Radius}

\begin{figure*}[b]
\plotfiddle{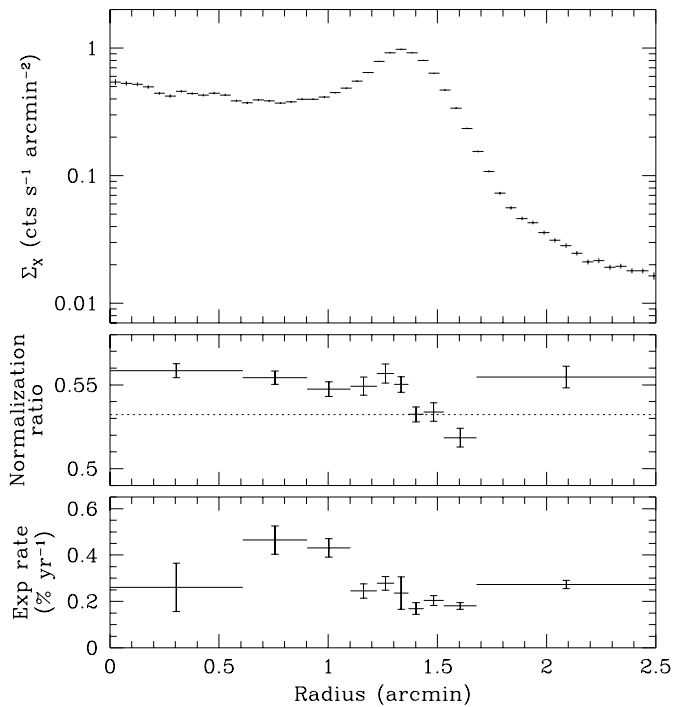}{3.6in}{0}{50}{50}{-150}{-75}
\figcaption[fig7.ps] {Radial X-ray surface brightness profile of
Kepler's supernova remnant from the \rosat\ HRI (top panel) and the
ratio of fitted normalizations (middle panel) and expansion rate as a
function of radius (bottom panel) from a comparison of the two \rosat\
HRI observations. The dashed line in the middle panel corresponds to
the ratio of detector live times. The radial bins for the expansion
measurements are not equally spaced in radius because the bin sizes
were chosen to contain roughly equal numbers of detected X-ray events
in each bin.
\label{fig7}}
\end{figure*}

How the expansion rate of Kepler's SNR varies with radius is also of
interest.  The procedure used here was the same as the one used to
determine the azimuthal expansion rates with the exception that,
instead of pie-shaped azimuthal regions, a number of disjoint
(non-overlapping) annular regions centered on the position of the
expansion center defined the separate regions over which individual
fits were done. The annuli were not uniformly spaced in radius, but
rather their various sizes were chosen so that each region contained
roughly 20,000 X-ray events.  The annular widths varied from a minimum
of $4^{\prime\prime}$ right at the bright limb of Kepler's SNR to a
maximum of $\sim$40$^{\prime\prime}$ for the first radial bin in the
fainter interior.  It is not surprising that the radial expansion
rates are sensitive to the relative registration of the two \rosat\
images. For the results quoted here I used the best-fit registration
from the previous azimuthal expansion study (i.e., corresponding to
the one with no significant sinusoidal term). The expansion rates are
shown in Figure 7 (bottom panel), along with, for reference, the X-ray
surface brightness profile (top panel) and the ratios of fitted
normalizations (middle).

The average of the radial rates plotted in fig.~7 weighted by their
respective uncertainties yields a mean expansion rate of $\sim$0.23\%
yr$^{-1}$, which is entirely consistent with the global mean expansion
rates determined previously (in both \S3.1 and \S3.2).  Near the
bright rim of the SNR, where the sensitivity to an expansion signal
should be greatest, the radial expansion rates are in the range
0.17--0.28\% yr$^{-1}$, with some indication that the rates are
slightly higher on the inner edge of the rim, compared to the outer
edge.  Over some parts of the remnant interior the rates are
remarkably high, a result of the apparent motion of the X-ray emitting
blobs in this area.  In particular the high rates for the second and
third radial bin shown in fig.~7 are due to the rapidly moving knot at
position angle 240$^\circ$ pointed out earlier (in \S3.2, also see
fig.~6) that straddles both annular regions. The rates in the interior
are also most sensitive to the choice of expansion center and the
implicit assumption of purely radial motion.  For example, if the
expansion center is displaced by the same $2\myarcsec5$ shift used
previously, the largest change in fitted rates occurs for the second
radial bin which drops by 2.7 standard deviations, while the other
points, particularly those near the bright rim, change much less.

The ratio of normalizations follows a very clear pattern.  The results
plotted in fig.~7, taken at face value, indicate that most of the
remnant has decreased in X-ray surface brightness by about 4\% over
the time interval between RHRI observations.  The outer edge of the
remnant shell, however, has remained at constant brightness or has
even increased slightly. These changes at the edge are statistically
significant and are in addition to any changes in brightness caused by
the remnant's expansion.  Note that the increase in overall surface
area of the remnant, due to its expansion, compensates for most of the
decrease in surface brightness so that the total count rate decreases
by only about 1\% between the two epochs, as mentioned earlier.
Although the overall uncertainty on the temporal variation of the RHRI
efficiency is of order a few percent (\S 2.1), it seems highly
unlikely that the dramatic spatial variation in brightness seen here
could be a result of instrumental changes.

\section{Discussion}

The extensive analysis described above is intended to convince the
reader that a robust measurement of the expansion rate of the X-ray
remnant of Kepler's SN has been made.  This confidence is based on 
deriving consistent values for the global mean expansion rate from
fits
\begin{itemize}
\item over the entire X-ray image of the remnant for five different
time baselines using high resolution images from both \rosat\ and
\einstein;
\item  as a function of azimuthal angle after correction for a
relative registration error ($<$1$^{\prime\prime}$) between the two
\rosat\ observations; and 
\item as a function of radius for optimal relative position registration
between the two \rosat\ observations.
\end{itemize}
As indicated, the latter two methods use only the two \rosat\
observations since they are the most sensitive and least susceptable
to systematic errors. These different techniques all produce rates
that are fully consistent with a value of $0.239^{+0.015\,+0.017}
_{-0.015\,-0.010}\,\%\, \rm yr^{-1}$, which accounts for both statistical
and systematic uncertainties.  This result can also be expressed as
an expansion time scale: $418^{+28\,+19}_{-24\,-27}\, \rm yr$.  Kepler's SNR is
the remnant of a historical SN and so the age is known: it was 393 yr
old in 1997.  Comparison of the current expansion time scale to the
age of the remnant reveals the extraordinary nature of these results.
The X-ray remnant has appeared to undergone little or no deceleration
since exploding as a supernova in 1604.  Specifically, the expansion
parameter, is measured to be $m =0.93^{+0.06\,+0.07}_{-0.06\,-0.04}$.
In the following this result is examined in the
context of other observational results as well as theoretical studies
of the dynamics of Kepler's SNR.

Optical knots are expanding quite slowly in Kepler's SNR. Bandiera \&
van den Bergh (1991) determine an expansion time scale of 32,000 $\pm$
12,000 yrs from the measured proper motions of some 50 knots of
optical nebulosity, most of which are located toward the northwestern
portion of the remnant.  This corresponds to an expansion rate of
$\sim$0.31 \% century$^{-1}$, which is nearly two orders of magnitude
smaller than the X-ray expansion rate.  There is also a translation of
the expansion center of the optical knots toward the northwest at a
rate of $0\myarcsec8$ century$^{-1}$.  The optical knots are believed
to be circumstellar in origin, so their expansion would therefore
largely arise from the motion of the original wind of the putative
massive-star progenitor, plus some momentum imparted to them by the SN
shock as well.  The proper motion of the optical remnant itself, which
is directed away from the Galactic plane, is strong evidence that the
progenitor was a high velocity ($\sim$280 km s$^{-1}$) star that left
the Galactic plane some $4\times 10^6$ yr ago, as originally proposed
by Bandiera (1987). It is interesting to note that Bandiera \& van den
Bergh (1991) also find a number of optical knots (most of which lie
near the edge of the shell) that appear to have brightened quite
rapidly, with turn-on times of only a few years.  Estimating these
knots to be roughly 1$^{\prime\prime}$ in angular size and making the
arguable assumption that the SN blast wave must be moving fast enough
to fully engulf a cloud during its period of brightening, would
suggest a shock speed on the order of $0\myarcsec2$
yr$^{-1}$. According to the X-ray image, the shock front is located at
a radius of $\sim$100$^{\prime\prime}$, so this estimate of the shock
speed is broadly consistent with the high X-ray expansion rates.
Clearly this is not intended to be a precise estimate, but does
indicate a plausible level of agreement.

The X-ray expansion rates of Kepler's SNR are also much higher than
the measurements in the radio band. Dickel et al.~(1988) present a
study of the expansion of Kepler's SNR based on VLA radio data from
two epochs separated by a time interval of about 4 yr.  They find a
mean expansion parameter for the entire radio remnant of $m=0.50$, as
well as azimuthal variations that range from $m=0.35$ on the bright
northern rim to $m=0.65$ on the eastern edge. These results are in
considerable disagreement with the X-ray results, both in terms of the
mean expansion rate and its variation with azimuth. However, Kepler is
not the first remnant to show this dicrepancy: both Tycho's SNR and
Cas A are expanding more rapidly in the X-ray band than in the
radio. The expansion parameter for Tycho's SNR is $m=0.72\pm0.10$ in
the X-ray band (Hughes 1997) and $m\sim 0.47$ in the radio band
(Strom, Goss, \& Shaver 1982). Cas A has global expansion parameters
of $m=0.73$ (X-ray) and $m=0.35$ (radio) (Koralesky et al.~1998, Vink
et al.~1998).  For Cas A there is also an indication that the size of
the X-ray remnant is smaller than the radio remnant, although the size
difference has been decreasing with time as the X-ray remnant
catches up to the radio. A similar study comparing the sizes of the
X-ray and radio images of Kepler's SNR is beyond the scope of this
paper but will be pursued in future work.

Chevalier (1982) has derived similarity solutions for the evolution of
young SNRs that can be used to put the Kepler expansion results in
some context.  For ejecta with power-law density profiles
characterized by index $n$ expanding into a circumstellar medium (CSM)
with index $s$, the expansion of the remnant is given by $R\propto
t^{(n-3)/(n-s)}$.  For a stellar wind profile ($s=2$), the measured
mean X-ray expansion rate for Kepler would suggest that the ejecta
have a steep density distribution, $n \gsim 11$, similar to the
atmosphere of a red supergiant. (A uniform density CSM, $s=0$ would
require a considerably steeper density distribution for the ejecta
profile to explain the mean X-ray expansion rate.)  In order to
explain the azimuthal X-ray brightness variations requires a
difference in density from the northwest to the southeast (all other
things being equal) of at least a factor of 3.  In the Chevalier model
the radius depends on the density of the stellar wind as $R\propto
(1/q)^{1/(n-s)}$, so that if the density varies by a factor of 3, the
radius should vary by a factor of $3^{-1/(n-s)}$.  For $n>11$, then
the remnant's radius will vary by a factor between 0.89 and 1, and in
other words, should appear quite round, as in fact Kepler's SNR does.

The long-standing problem with this simple model has been in
understanding the origin of the stellar wind's strong azimuthal
asymmetry.  Bandiera (1987) realized that the interaction between the
ambient interstellar medium (ISM) and the stellar wind from the
rapidly-moving progenitor of Kepler's SN would naturally produce an
asymmetric CSM.  This would consist of a dense bow-shock shell ahead
of the star and an $r^{-2}$ stellar wind profile elsewhere, i.e., both
within the shell and in the direction away from the star's motion. The
asymmetry of the observed remnant today reflects the interaction of
the SN blast wave with the asymmetric CSM established by this process.
Borkowski et al.~(1992) examined this scenario in some detail and
concluded that ``the bow-shock model of Bandiera (1987) is currently
the only model which accounts for most of the properties of this
remnant.'' The slow expansion of the radio remnant as a whole ($m\sim
0.50$) and in particular the very small expansion rate toward the
northwest were considered strong evidence in support of the bow-shock
shell model. However, this model should be re-evaluated in light of
the much higher X-ray expansion results presented here. Furthermore
it now appears to be the case for Kepler, as was found earlier for both
Cas A and Tycho, that distinctly different magnetohydrodynamical
structures are giving rise to the different radio and X-ray emissions.

\section{Summary}

In this article I have presented the first measurement of the
expansion of Kepler's SNR in the X-ray band. The remnant is expanding
at a current rate averaged over the remnant of $0.239^{+0.015\,+0.017} 
_{-0.015\,-0.010}\,\%\, \rm yr^{-1}$, which is two orders of magnitude
faster than the expansion of the optical knots and roughly twice as
fast as the expansion of the radio remnant.  Kepler is the remnant of an
historical SN so the current rate can be compared to the time-averaged
rate to indicate how much deceleration has occurred.  Remarkably, the
current and time-averaged rates are nearly equal implying that the
remnant has hardly decelerated and is close to expanding freely.  The
expansion parameter of Kepler's SNR is $m
=0.93^{+0.06\,+0.07}_{-0.06\,-0.04}$, including both statistical and
systematic uncertainty (the first and second set of
errors, respectively).

There is significant variation of the expansion rate as a function of
position in the SNR, although much of it does not appear to be
correlated with obvious emission features.  Of particular note is the
finding that the faint, southeastern part of the remnant is expanding
at roughly the same rate as the bright, northwestern part. The most
rapidly moving feature in Kepler's SNR is a knot of X-ray emission
about two-thirds of the way out from the center in the southwestern
quadrant, which has an expansion timescale of $\sim$250 yr, only a
fraction of the remnant's age.  It is not clear what the origin of
this motion is.

One might be tempted to use the X-ray expansion rate measured here and
the shock velocity (1550--2000 km s$^{-1}$) determined from the
properties of the H$\alpha$ emission (Blair et al.~1991) to estimate a
distance to Kepler's SNR.  I resist this temptation, since it relies
on a number of assumptions that are unlikely to be correct. Moreover,
within the next year it will become possible, using \chandra, \xmm, or
\astroe, to measure the Doppler shifts from bright X-ray emission
lines of individual knots seen in projection near the center of the
remnant. Combined with the X-ray expansion rate, this will provide a
direct and independent measurement of the distance to Kepler's SNR.
To be strictly correct one will need to carry out the angular
expansion measurement in the same emission line as the Doppler
velocity mapping is done since chemical stratification of
homologously-expanding ejecta would produce different expansion rates
for different species.

The current best model for the evolution of Kepler's SNR was developed
to explain two important observations: the proper motion of the
progenitor star and the high density of the ambient environment.  This
model has been shown to be consistent with a number of other
observations as well, including the present slow expansion rate of the
radio remnant.  At first glance therefore it would appear that the
extraordinarily rapid expansion rate of the X-ray remnant found here
should be grossly inconsistent with these favored dynamical models.
Future work should focus on this by identifying the
magnetohydrodynamical structures, e.g., reverse shock, blast wave,
reflected or transmitted shock, etc., that correspond to the X-ray and
radio emission by modeling their dynamical and radiative processes.

Kepler is now the third young X-ray SNR to have its expansion rate
measured; Cas A and Tycho are the others. In each of these three cases
the same general effect is seen.  The radio remnant and the X-ray
remnant are of the same size, so the time-averaged expansion rates are
quite similar.  However the current expansion rate of the X-ray
remnant is significantly larger than the current expansion rate of the
radio remnant, clearly indicating that the structures in the remnant
giving rise to the X-ray emission have been decelerated considerably
less than the radio-emitting structures.  Although perplexing, these
differences in dynamical evolution are basic enough that they surely
contain the seeds of some fundamental new insights into the nature and
evolution of young supernova remnants.

\acknowledgments

I would like to thank R.~Bandiera, L.~David, A.~Decourchelle,
M.~Gagn\'e, and S.~Snowden for discussions and help with various
aspects of this project. I thank John Dickel for his useful comments
as referee.  This research has made use of data obtained through the
High Energy Astrophysics Science Archive Research Center Online
Service, provided by the NASA/Goddard Space Flight Center.  Partial
support was provided by NASA grants NAG5-4794 and NAG5-6420.

\end{document}